\documentclass[twocolumn,prd,preprintnumbers,showpacs]{revtex4}

\usepackage{graphicx}% Include figure files
\usepackage{bm}% bold math
\usepackage{amssymb,amsmath}

\usepackage{array}

\def\beqn{\begin{eqnarray}}
\def\eeqn{\end{eqnarray}}
\def\barr{\begin{array}}
\def\earr{\end{array}}
\def\btab{\begin{tabular}}
\def\etab{\end{tabular}}
\def\bite{\begin{itemize}}
\def\eite{\end{itemize}}
\def\bcen{\begin{center}}
\def\ecen{\end{center}}

\def\eq{\begin{equation}}
\def\ee{\end{equation}}
\def\nn{\nonumber}

\def\kdagger{K\hspace{-0.22cm}/}

\def\pgdagger{P\hspace{-0.22cm}/\,}

\def\pdagger{p\hspace{-0.19cm}/}

\def\keldagger{k\hspace{-0.2cm}/}

\def\q2dagger{q_2\hspace{-0.35cm}/\;}

\begin{document}

%%%%%%%%%%%%%%%%%%%%%%%%%%%%%%%%%%%%%%%%%%%%%%%%%%%%%%%
%%%%%%%%%%%%%%%%%%%%%%%%%%%%%%%%%%%%%%%%%%%%%%%%%%%%%%%

%\preprint{MITP/14-039}
\title{Forward sum rule for the $2\gamma$-exchange correction to the charge radius extraction from elastic electron scattering}

\author{Mikhail Gorchtein} 
\affiliation{
Institut f\"ur Kernphysik, Universit\"at Mainz, 55128 Mainz, Germany} 

\date{\today}

\begin{abstract}
Two-photon exchange (TPE) contributions to elastic electron-proton scattering in the forward regime in leading logarithmic $\sim t\ln|t|$ approximation in the momentum transfer $t$ are considered. The imaginary part of the TPE amplitude in the forward kinematics is related to the total photoabsorption cross section. The real part of the TPE amplitude is obtained from an unsubtracted fixed-$t$ dispersion relation. This allows for a clean prediction for the real part of the TPE amplitude at forward angles with the leading term $\sim t\ln|t|$. Numerical estimates are comparable with or exceed the experimental precision in extracting the charge radius from the experimental data. 
\end{abstract}


\maketitle

%%%%%%%%%%%%%%%%%%%%%%%%%%%%%%%%%%%%%%%%%%%%%%%%%%%%%%%%%%%%%%%%%%%%%%
%%%%%%%%%%%%%%%%%%%%%%%%%%%%%%%%%%%%%%%%%%%%%%%%%%%%%%%%%%%%%%%%%%%%%%
%\section{Introduction}
%\label{sec:intro}
Nucleon structure has been studied with elastic electron scattering since the 1950's. By means of the Rosenbluth separation the measurement of the unpolarized cross section allows to extract the electromagnetic form factors of the nucleon. 
The interest in measuring the elastic cross section at low (negative) $t$ is, e.g., the extraction of the slope of the electric Sachs form factor $G_E$ that is related to the charge radius $R_{E}$ as 
\beqn
G_E(t\to0)=1+R^2_E t/6+O(t^2)
\eeqn

A recent measurement at Mainz \cite{Bernauer:2010wm} led to the most precise ($\lesssim1\%$) determination of the proton charge radius with electron scattering experiments to date,
\beqn
R^p_{E}=0.879\pm0.008\,{\rm fm},
\eeqn
where the uncertainty quoted above represents a combined statistical, systematical, model-dependent and group-dependent uncertainties defined in that Ref. Proton charge radius is extracted from hydrogen spectroscopy data with even higher precision \cite{Mohr:2012tt}
\beqn
R^p_{E}=0.8775\pm0.0051\,{\rm fm},
\eeqn
the two methods delivering results that are in a very nice agreement. The recent Lamb shift measurements in muonic hydrogen 
\cite{Pohl:2010zza,Antognini:1900ns} lead to an extraction of the proton charge radius that is ten times more precise,
\beqn
R^p_{E}=0.84087\pm0.00039\,{\rm fm},
\eeqn
and differs by seven standard deviations from the value obtained with electronic hydrogen and in scattering experiments. 
In the context of the  ``proton radius puzzle", as this discrepancy was coined in the literature, nucleon structure-dependent corrections to the Lamb shift, most notably the two-photon exchange (TPE) correction, underwent a renewed scrutiny with two methods that provide a controlled estimate of the systematical uncertainty of such a calculation: the dispersion relations \cite{Pachucki:1999zza,Carlson:2011zd,Gorchtein:2013yga} and within effective theories \cite{Nevado:2007dd,Hill:2011wy,Birse:2012eb,Alarcon:2013cba}; however the discrepancy is still present. For electron scattering, dispersion relations have the potential to provide model-independent calculations of the TPE effect \cite{Borisyuk:2006uq,Borisyuk:2008es,Tomalak:2014dja}, although these references only account for the ground state contribution to TPE. I refer the reader to a recent review of the TPE effects in electron scattering \cite{Arrington:2011dn}.

%In this work, we revisit the two-photon exchange correction to the elastic electron scattering at low momentum transfer. The article is organized as follows. In Section II, kinematics and cross section for elastic $ep$-scattering are considered in the low-$t$ regime. Section III introduces the TPE effects and considers the modifications of the cross section in their presence. In Section IV, the imaginary part of the (nearly) forward TPE amplitude is obtained, and the sum rule for the term $t \ln |t|$ due to TPE is obtained in terms of an energy-weighted integral over the total photo absorption cross section. Numerical results are presented in Section V, and their consequences for the existing and upcoming experimental data are discussed. A short conclusion closes the article.

%\section{Elastic $ep$-scattering amplitude}
%\label{sec:el_ampl}
\vspace{0.25cm}
In this work, I reexamine the two-photon exchange correction to elastic electron scattering at low momentum transfer. I consider elastic electron-proton scattering process $e(k)+p(p)\to e(k')+p(p')$ for which I define $P=(p+p')/2$, $K=(k+k')/2$, $\Delta=k-k'=p'-p$ and choose the invariants $t=\Delta^2=-Q^2<0$ and $\nu=(P\cdot K)/M$ as the independent variables, where $M$ denotes the nucleon mass, and the electron mass $m_e$ is neglected. They are related to the Mandelstam variables $s=(p+k)^2$ and $u=(p-k')^2$ through $s-u=4M\nu$ and  $s+u+t=2M^2$. The usual polarization parameter $\varepsilon$ is related to the invariants $\nu$ and $t$ as
\beqn
\varepsilon\,=\,\frac{\nu^2-M^2\tau(1+\tau)}{\nu^2+M^2\tau(1+\tau)},
\eeqn
with $\tau=-t/(4M^2)$. Elastic scattering of a massless electron off a spin-$1/2$ target in the Born (one photon exchange, OPE) approximation is described by the familiar Dirac and Pauli form factors $F_{1}$ and $F_{2}$, respectively,
\beqn
T_{B}=\frac{e^2}{-t}
\bar{u}(k')\gamma_\mu u(k)\,
\bar{u}(p')
\left[{F}_1 \gamma^\mu+
{F}_2\frac{i\sigma^{\mu\alpha}\Delta_\alpha}{2M}\right]u(p).\label{Born}
\eeqn
\indent
The unpolarized cross section is 
\beqn
\frac{d\sigma}{d\Omega_{Lab}}&=&\frac{4\alpha^2\cos^2\frac{\Theta}{2}}{t^2}
\frac{E'^3}{E}\sigma_R,
\eeqn
with $\Theta$ the electron Lab scattering angle and $E(E')$ the incoming (outgoing) 
electron Lab energy. The reduced cross section $\sigma_R$ is expressed in terms of electric and magnetic Sachs form factors $G_E=F_1-\tau F_2$ and $G_M=F_1+F_2$, respectively, as
\beqn
\sigma_R=[G_E^2+\frac{\tau}{\varepsilon} G_M^2]/(1+\tau).
\eeqn

Before going on to discuss the two-photon exchange I wish to determine the level of accuracy that modern experiments set for this calculation. To this end, the reduced cross section taken in Born approximation can be expanded in a Taylor series in powers of negative $t$. Keeping the linear terms in this expansion, I write
\beqn
\sigma_R^B=1+\frac{1}{3}R_E^2 t-\frac{t\mu_p^2}{4M^2\varepsilon}+\frac{t}{4M^2}+O(t^2),\label{eq:sigmaRB}
\eeqn
with $\mu_p=G_M^p(0)\approx2.793$ the proton magnetic moment in units of the nuclear magneton. Correspondingly, the 1\% relative uncertainty in the charge radius is translated into the uncertainty in the reduced cross section 
\beqn
\delta\sigma_R^B=\frac{1}{3}R_E^2 \frac{2\delta R_E}{R_E}|t|\approx0.120\frac{|t|}{{\rm GeV}^2}.
\eeqn

For the smallest values of $|t|$ accessed in the A1 experiment, $|t|_{min}=4\times10^{-3}$ GeV$^2$ the relative uncertainty of $\sigma_R$ is of order $5\times10^{-4}$, similar to the natural size of the order $\alpha_{em}$ correction, $\sim\alpha_{em}/(4\pi)$. Most order $\alpha_{em}$ corrections can be calculated quite reliably, the exception being the two-photon exchange. The latter is included approximating the TPE graph by only the ground state contribution that is furthermore approximated according to Mo and Tsai \cite{Mo:1968cg} or Maximon and Tjon \cite{Maximon:2000hm}, as well as the so-called Feshbach correction \cite{McKinley:1948zz}, leading to a generic result $\delta\sigma_R^{OPE+R.C.}=(1+\delta)\sigma_R^B$ with the correction $\delta\sim\alpha_{em}$.  We discuss the two corrections in more detail in the following section. An inclusion of the general nucleon structure in the TPE is complicated and is only possible in forward kinematics, as e.g. in the calculation of the polarizability correction to the Lamb shift. It is possible to show that such inelastic contributions should vanish for $t=0$, but can lead to the behavior $t\ln |t|$. This behavior was obtained in Ref. \cite{Gorchtein:2006mq} that concentrated on high energy regime. Terms $\sim t\ln |t|$ introduce a substantial nonlinearity of the reduced cross section as function of $t$ at low $t$, the opposite to the OPE contribution in Eq. (\ref{eq:sigmaRB}) that becomes more linear at lower $t$. Present work is dedicated to assessing this correction in the kinematics of the relevant experiments, from a few hundred MeV to a few GeV beam energy and $|t|\lesssim0.1$ GeV$^2$. The same approach is expected to be relevant for the measurement of the deuteron charge radius in elastic $eD$-scattering, and the respective estimates will also be presented. 

%\section{Two-photon exchange amplitude}
%\label{sec:TPE}
\vspace{0.25cm}
In the presence of the TPE effects, and in the approximation of small electron mass, the elastic $ep$-scattering amplitude is given by three scalar amplitudes $\tilde F_i(\nu,t)$,
\beqn
T&=&\frac{e^2}{-t}
\bar{u}(k')\gamma_\mu u(k)\label{f1-3}
\\
&\times&
\bar{u}(p')
\left[\tilde{F}_1 \gamma^\mu+
\tilde{F}_2\frac{i\sigma^{\mu\alpha}\Delta_\alpha}{2M}+
\tilde{F}_3\frac{\kdagger P^\mu}{M^2}\right]u(p),\nn
\eeqn
In the one-photon exchange (OPE) approximation, the known Dirac and Pauli form factors are recovered,
%\beqn
$\tilde{F}_{1,2}^{OPE}(\nu,t)=F_{1,2}(t)$, while the third structure is absent, 
$\tilde{F}_3^{OPE}=0$.
%\eeqn
%\indent
I separate the TPE effects explicitly, 
\beqn
\tilde{F}_{1,2}&=&F_{1,2}+\delta\tilde{F}_{1,2},\nn\\
\tilde{G}_{E,M}&=&G_{E,M}+\delta\tilde G_{E,M},
\eeqn
where the generalization of the Sachs form factors was introduced, $\tilde G_E=\tilde F_1-\tau\tilde F_2$ and $\tilde G_M=\tilde F_1+\tilde F_2$. In presence
of TPE effects, reduced cross section $\sigma_R$ reads
\beqn
\sigma_R&=&\frac{G_E^2+{\tau} G_M^2/{\varepsilon}}{1+\tau}
+\frac{2G_E}{1+\tau} {\rm Re}
\left(\delta\tilde{G}_E+\frac{\nu}{M}\tilde{F}_3\right) \nn\\
&+&2 \frac{\tau}{\varepsilon(1+\tau)} G_M {\rm Re}
\left(\delta\tilde{G}_M+\varepsilon\frac{\nu}{M}\tilde{F}_3\right)\\
&=&\frac{G_E^2+{\tau} G_M^2/{\varepsilon}}{1+\tau}
+\frac{2G_E}{1+\tau} {\rm Re}
\left(\!\delta\tilde{G}_E+\frac{\nu}{M}\tilde{F}_3\!\right)+{\cal{O}}(\alpha
t^2)\nn.% +{\cal{O}}(e^4)
\eeqn

%We recall that at low $|t|$ the inelastic contribution to TPE should go as $\alphat$. 

%To proceed in a systematic way, we will only keep leading order in
%$t$ and neglect all terms that are of order $\alpha t^2$ and higher. 
%This leads to 
%\beqn
%\sigma_R=\frac{G_E^2+\frac{\tau}{\varepsilon} G_M^2}{1+\tau}
%+ 2 {\rm
%  Re}\left[\delta\tilde{F}_1+\frac{E}{M}\tilde{F}_3\right]+{\cal{O}}(\alpha
%t^2),
%\eeqn
%where the LAB energy $E=\frac{s-M^2}{2M}$, the $t=0$ limit of $\nu=\frac{s-M^2+t/2}{2M}$ was introduced. 

It is straightforward to see that the TPE effect on the unpolarized cross section at low $t$ depends on the same combination of the amplitudes %, $\delta\tilde{G}_E(\nu,t)+(\nu/M)\tilde{F}_3(\nu,t)\equiv\Phi(\nu,t)$, 
as the elastic amplitude averaged over nucleon spins,
\beqn
&&\bar T_{2\gamma}=\frac{e^2}{-t}
\bar{u}(k')\gamma_\mu u(k)\label{eq:Phi}\\
&&\times\frac{{\rm Tr}(\pdagger'+M)\left[\delta\tilde{F}_1 \gamma^\mu+
\delta\tilde{F}_2\frac{i\sigma^{\mu\alpha}\Delta_\alpha}{2M}+
\tilde{F}_3\frac{\kdagger \;P^\mu}{M^2}\right](\pdagger+M)}{8M}\nn\\
&&=\frac{e^2}{-t} \bar{u}(k')\pgdagger u(k)\!\!\left[\delta\tilde{G}_E+\frac{\nu}{M}\tilde{F}_3\right]\!\!\equiv
\frac{e^2}{-t} \bar{u}(k')\pgdagger u(k)\Phi(\nu,t).\nn
\eeqn
%where upon keeping leading terms in $t$ only a short hand was introduced, 
%\beqn
%\Phi(\nu,t)=\delta\tilde{G}_E(\nu,t)+\frac{\nu}{M}\tilde{F}_3(\nu,t)
%\eeqn

%\section{Imaginary part of the near-forward elastic $ep$-scattering amplitude}
%\section{Near-forward elastic $ep$-scattering amplitude from a dispersion relation}
%\label{sec:im_2gamma}
\begin{figure}[h]
{\includegraphics[height=2cm]{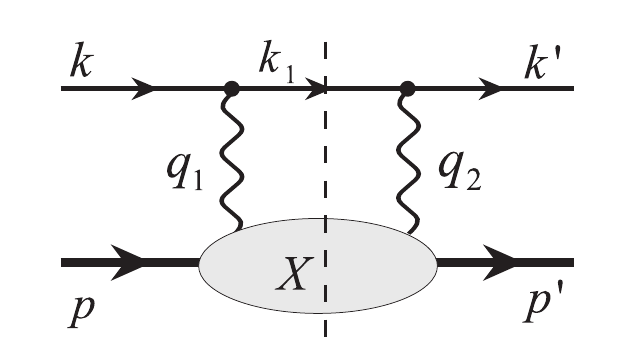}}
\caption{Imaginary part of the $2\gamma$-exchange diagram}
\label{fig:2gammadiag}
%\vspace{-1cm}
\end{figure}
The imaginary part of the TPE diagram in Fig. \ref{fig:2gammadiag} 
is given by the phase-space integral
\beqn
2{\rm Im}T_{2\gamma}&=&e^4\int\frac{d^3\vec{k}_1}{(2\pi)^32E_1}
\frac{\ell_{\mu\nu}\cdot{\rm Im}W^{\mu\nu}}{(q_1^2+i\epsilon)(q_2^2+i\epsilon)},
\eeqn
where the leptonic tensor is given by 
\beqn
\ell_{\mu\nu}=\bar{u}(k')\gamma_\nu(\keldagger_1+m_e)\gamma_\mu u(k)\approx\bar{u}(k')\gamma_\nu\keldagger_1\gamma_\mu u(k),\label{eq:lmunu}
\eeqn
and the on-shell condition for the intermediate electron leads to $E_1=(\vec k_1^2+m_e^2)^{1/2}\approx|\vec k_1|$.
The hadronic tensor can be split into elastic and inelastic contributions, $W^{\mu\nu}=W^{\mu\nu}_{el}+W^{\mu\nu}_{inel}$. This separation is possible because the former has a pole, Im$W_{el}\sim\delta((p+q_1)^2-M^2)$, whereas the latter has a unitarity cut starting at the pion production threshold $(p+q_1)^2=(M+m_\pi)^2$.

%\subsection{Elastic contribution}
The imaginary part of the elastic part is due to the on-shell nucleon in the intermediate state,
\beqn
{\rm Im}W^{\mu\nu}_{el}&=&
2\pi\delta((P+K-k_1)^2-M^2)\label{eq:wmunu-el}\\
&\times&\bar{u}(p')\Gamma^{*\nu}(q_2)(\pgdagger+\kdagger-\keldagger_1+M)\Gamma^\mu(q_1) u(p),\nn
\eeqn
with $\Gamma^\mu(\Delta)=F_1(\Delta^2)\gamma^\mu+F_2(\Delta^2)i\sigma^{\mu\alpha}\Delta_\alpha/(2M)$ the on-shell nucleon electromagnetic vertex. 
It contains the infrared (IR) divergent part that is logarithmic in the fictitious photon mass, $\sim\ln\lambda^2$, the coefficient in front of it is model-independent, and has been calculated in Refs. \cite{Mo:1968cg,Maximon:2000hm} using the soft photon approximation in the loop. Ref. \cite{Mo:1968cg} used the approximation $q_1\approx0,\,q_2\approx\Delta$ and vice versa both in the numerator and the denominator of the integral, the result simply factorizing the one-photon exchange (Born) amplitude as 
\beqn
{\rm Im} \Phi^{(a)}=-\frac{\alpha E_{cm}^2}{\pi }G_E(t)\!\!\int\!\frac{d\Omega_1}{q_1^2}
=\alpha\ln\!\left(\!\!\frac{4E_{cm}^2}{\lambda^2}\!\!\right)\!G_E(t),\;\label{eq:mo-tsai}
\eeqn
with c.m. energy of the electron $E_{cm}\approx(s-M^2)/2\sqrt s$, neglecting the electron mass. On the other hand, Ref. \cite{Maximon:2000hm} applied the soft photon approximation in the numerator only leading to 
\beqn
{\rm Im} \Phi^{(b)}=\frac{-\alpha tE_{cm}^2}{2\pi}G_E(t)\!\!\int \!\!\frac{d\Omega_1}{q_1^2q_2^2}
=\alpha\ln
%\left[
\frac{-t}{\lambda^2}
%\right]
G_E(t).\label{eq:maximon-tjon}
\eeqn
%The limit of small electron mass was taken in the above results. 
The real part is obtained from a dispersion relation at fixed $t$, 
\beqn
{\rm Re} \Phi(\nu,t)&=&\frac{2\nu}{\pi}{\cal P} \int_{\nu_0^{el}}^\infty\frac{d\nu'}{\nu'^2-\nu^2}{\rm Im} \Phi(\nu',t),
\label{eq:PhiDR}
\eeqn
with $\nu_0^{el}=t/(4M)\leq0$ the threshold for the $s$-channel unitarity cut. The evaluation of the dispersion integral with the imaginary part of Eq. (\ref{eq:maximon-tjon}) yields %for the real part
\beqn
{\rm Re} \Phi^{(b)}&=&\frac{\alpha}{\pi}\ln\left(\frac{-t}{\lambda^2}\right)G_E(t)\ln\left(\frac{4M\nu+t}{4M\nu-t}\right).\label{eq:soft}
%\nn\\&\approx&\frac{\alpha}{\pi}G_E(t)\frac{t}{2ME}\ln\left(\frac{-t}{\lambda^2}\right),\label{eq:soft}
\eeqn
%where the LAB energy $E=(s-M^2)/2M$, the $t=0$ limit of $\nu=(s-M^2+t/2)/2M$ was introduced.

While the imaginary part of $\Phi^{(b)}$ behaves as $\ln(-t/\lambda^2)$, its real part is suppressed by an extra power of $t$ coming from the second logarithm. 
%This is a consequence of the principal value integral vanishing identically at $t=0$.
%\beqn
%{\cal P}\int_{t/(4M)}^\infty\frac{2\nu d\nu'}{\nu'^2-\nu^2}=\ln\left(\frac{4M\nu+t}{4M\nu-t}\right)=O(t).
%\eeqn

The result of Eq. (\ref{eq:soft}) was used in the analysis of the low-$t$ data from Mainz \cite{Bernauer:2010wm} (without the low-$t$ approximation), and I use the IR part of the TPE amplitude in this form to define the IR finite part of the elastic box as 
\beqn
\Phi^{el}_{F}\equiv\Phi^{el}-\Phi^{(b)},
\eeqn
that should be added to the full set of radiative corrections included in the experimental analysis. 
A straightforward calculation using the hadronic tensor of Eq. (\ref{eq:wmunu-el}), the leptonic tensor of Eq. (\ref{eq:lmunu}), and the definition of the amplitude $\Phi$ in Eq. (\ref{eq:Phi}), I obtain
%\begin{widetext}
\beqn
&&{\rm Im} \Phi^{el}_{F}(\nu,t)=\frac{-\alpha tE_{cm}^2}{2\pi}\int\frac{d\Omega_1}{q_1^2q_2^2}\nn\\
&&\times\left\{F_{11}F_{12}-F_1+\frac{q_1^2F_{12}F_{21}+q_2^2F_{11}F_{22}-tF_2}{4M^2}\right.\nn\\
&&+\left.\frac{t-q_1^2-q_2^2}{8M^2}\left[\mu_p^2-1-\frac{4sM^2}{(s-M^2)^2+st}\right]\right\},
\eeqn
%\end{widetext}
where terms that cannot lead to $t\ln t$-behavior were dropped. For compactness, the shorthand $F_{ij}=F_i(q_{j}^2)$ and $F_i=F_i(t)$ was introduced. The above integral is IR finite as it depends on one master integral over the solid angle of the intermediate electron,
\beqn
\int d\Omega_1\frac{t-q_1^2-q_2^2}{q_1^2q_2^2}=\frac{2\pi}{E_{cm}^2}\ln\left(\frac{4E_{cm}^2}{-t}\right).
\eeqn 
\indent
Expanding the form factors under the integral as $F_i(q^2)=F_i(0)+q^2 F_i'(0)+\dots$, we obtain %for Im$\Phi$
%\begin{widetext}
\beqn
{\rm Im} \Phi^{el}_{F}=\frac{\alpha t}{2}\ln\!\left(\!\frac{4E_{cm}^2}{-t}\!\right)\!\!\left[\frac{1}{4E_{cm}^2+t}+\frac{R_E^2}{3}-\frac{\mu_p^2-1}{4M^2}\right]\!\!,
\eeqn
where I used the relation $F_1'(0)=R_E^2/6-F_2(0)/(4M^2)$. 
%\end{widetext}
%This result is further simplified since terms that are constant in energy, as e.g. $t\ln t$ will be suppressed by an extra power of $t$ when calculating the principal value integral, as discussed above, while the other part of the logarithm will not lead to a $t\ln t$ behavior. Therefore, the final {\it exact} result for the leading log term $\sim t\ln t$ is independent of the details of the nucleon form factors and reads
%\beqn
%{\rm Im} \Phi^{el}_{hard}(\nu,t)=\frac{\alpha}{2} \frac{st}{(s-M^2)^2+st}\ln\frac{(s-M^2)^2}{-st}
%\eeqn
The real part is obtained according to Eq. (\ref{eq:PhiDR}), 
\beqn
{\rm Re} \Phi^{el}_{F}(\nu,t)\approx\frac{\alpha\pi}{2}\!\!\left[\frac{\sqrt{-t}}{2\nu+\sqrt{-t}}+t\left(\frac{R_E^2}{3}-\frac{\mu_p^2-1}{4M^2}\right)\right]\!\!.
\eeqn
\indent
The first term in the square bracket is the well-known Feshbach correction \cite{McKinley:1948zz}. I see this calculation as a useful cross check for the method of isolating the leading $t$-behavior (the Feshbach term was also found in a similar manner in Refs. \cite{Borisyuk:2006uq,Arrington:2011dn}). The second term was recently found in Ref. \cite{Lee:2014uia} which however missed the third term. The missed term amounts to $-0.075$ fm$^2$ which is not small if compared to $R_E^2/3\approx0.255$ fm$^2$. Moreover, due to the approximations made the term $\sim t$ is not model-independent: other terms without the logarithmic behavior were omitted, but they would contribute at the same order. Therefore, while I support the statement of Ref. \cite{Lee:2014uia} that the Feshbach correction alone is not enough to warrant the precision of the charge radius extraction in Bernauer et al., inclusion of the correction $\sim \alpha\pi r_E^2$ is also not sufficient. The main message to take home from this exercise is that the method allows one to obtain the leading-$t$ behavior. In what follows I show that the term $\sim t\ln t$ that arises from the inelastic states in the box can be calculated model-independently, as well.
%An important lesson that one learned is that no terms $\sim t\ln t$ arise from the elastic box, only $\sim \sqrt{-t}$ and $\sim const.\cdot t$. In what follows I show that the term $\sim t\ln t$ does arise from the inelastic states in the box, and that the coefficient in front of that term can be calculated model-independently. 

%\subsection{Inelastic contribution}
%\vspace{0.5cm}
I turn to the inelastic contribution, and study first the two representative integrals over the solid angle, 
\beqn
I_1&=&\int\frac{d\Omega_1}{q_1^2q_2^2}=\frac{2\pi}{-t(E_{1cm})^2}
\ln\left(\frac{(E_{1cm})^2}{(E_{cm}-E_{1cm})^2}\frac{-t}{m_e^2}\right)\!,\nn\\
I_2&=&-\int \!d\Omega_1\!\left[\frac{1}{q_1^2}+\frac{1}{q_2^2}\right]=\frac{2\pi}{E_{cm}E_{1cm}}\ln\!\left(\!\frac{4E_{1cm}^2}{m_e^2}\!\right),
\label{eq:I1,2}
\eeqn
where the c.m. energy of the intermediate electron is distinct from the external electron energy, $E_{1cm}=(s-W^2)/2\sqrt s$, and the invariant mass squared of the intermediate hadronic system, $W^2=(p+q_1)^2$ lies above the pion production threshold, $W^2\geq W_\pi^2=(M+m_\pi)^2$. Due to this threshold, the IR divergence is absent. However, the collinear divergence (emission of an energetic real photon collinear to the electron line) would be possible if the electron were massless. Keeping the finite mass of the electron makes the individual integrals finite, but a potential chiral divergence is introduced. It cannot appear in the final result, and one should expect this logarithmic dependence on the electron mass to vanish. I will be looking for the leading $t$-behavior that is expected to be $\sim t\ln t$, and that behavior can only come from the integral $I_1$. I will keep the integral $I_2$ to cancel the $\ln m_e^2$ dependence but neglect terms $\sim t$.%, consistently with the rest of this work. 
 
The spin-averaged part of the hadronic tensor with real photons in general (non-forward) kinematics is expressed in terms of two scalar amplitudes $f_{1,2}(P\cdot q_1,t)$ \cite{Tarrach:1975tu}
%\begin{widetext}
\beqn
W^{\mu\nu}&=&f_1\Big[(q_1\cdot q_2)g^{\mu\nu}-q_1^\nu q_2^\mu\Big]\\
&+&f_2\Big[(P\cdot q_1)^2g^{\mu\nu}+(q_1\cdot q_2)P^\mu P^\nu\nn\\
&&-(P\cdot q_1)(P^\mu q_1^\nu+P^\nu q_2^\mu)\Big].\nn
\eeqn
\indent
%\beqn
%&&\ell_{\mu\nu}\cdot{\rm Im}W^{\mu\nu}=f_1(Q_1^2+Q_2^2)\bar{u}(k')\keldagger_1 u(k)\\
%&&+f_2\left\{[4(P\cdot K)(P\cdot q_1)-2(P\cdot q_1)^2-(q_1\cdot q_2)P^2]\bar{u}(k')\keldagger_1 u(k)
%+[2(P\cdot k_1)(q_1\cdot q_2)-(P\cdot q_1)(Q_1^2+Q_2^2)]\bar{u}(k')\pgdagger u(k)\right\}\nn
%\eeqn
%\end{widetext}
Making use of the relation 
\beqn
\bar{u}(k')\keldagger_1 u(k)=\frac{t\frac{(P\cdot k_1)}{(P\cdot K)}-q_1^2-q_2^2}{4(P\cdot K)}\bar{u}(k')\pgdagger u(k),
\eeqn
performing tensor contraction and consistently neglecting terms $\sim tq_{1,2}^2$ and $\sim q_1^2 q_2^2$ in the numerator, the imaginary part of the TPE amplitude can be cast in the form
\beqn
&&{\rm Im}T_{2\gamma}=e^4\bar{u}(k')\pgdagger u(k)\int\frac{d^3\vec{k}_1}{(2\pi)^32E_1}
\frac{1}{q_1^2q_2^2}\label{eq:imT2g_Imf2}\\
&&\times\frac{(P\cdot K)^2+(P\cdot k_1)^2}{2(P\cdot K)}\left[t\frac{(P\cdot k_1)}{(P\cdot K)}-q_1^2-q_2^2\right]{\rm Im} f_2,\nn
\eeqn
while the amplitude $f_1$ does not contribute at the leading logarithm accuracy. According to the power counting used throughout this calculation, Im$\,f_2(P\cdot q_1,t,q_1^2,q_2^2)$ should be taken at $t=q_1^2=q_2^2=0$. In these kinematics, the optical theorem relates this imaginary part to the total real photoabsorption cross section $\sigma_T$ as
\beqn
{\rm Im}f_2(P\cdot q_1,0,0,0)=-2\sigma_T/[(P\cdot q_1)e^2].
\eeqn
\indent
Using the definition of Eq. (\ref{eq:Phi}) and identifying the solid angle integrals in Eq. (\ref{eq:imT2g_Imf2}) with the previously introduced $I_{1,2}$ in Eq. (\ref{eq:I1,2}), one can express the leading logarithm contribution to the imaginary part of the elastic $ep$-scattering amplitude near the forward direction as
\beqn
{\rm Im}\Phi=\frac{-t}{4\pi^2}\!\int\limits_{E_\pi}^E\!\frac{d\omega}{\omega}\sigma_T(\omega)\ln\!\left(\frac{4\omega_{cm}^2}{-t}\right)\!
\left[1-\frac{\omega}{E}+\frac{\omega^2}{2E^2}\right]\!,\;
\eeqn
with $\omega=(W^2-M^2)/2M$ the LAB real photon energy, and $\omega_{cm}=M\omega/\sqrt s$ the c.m. photon energy.
The dispersion integral starts from the pion threshold, %at fixed $t$ \cite{Gorchtein:2006mq},%\footnote{The reader may note that the amplitude $\Phi$ is an odd function of energy, quite opposite to the Lamb shift calculation where the TPE amplitude is an even function of energy. These two facts are not in contradiction since there are indeed two different amplitudes for a scattering of an electron off a spinless target: $\Phi\bar u\pgdagger u$ and $\phi m_e\bar u u$. The latter will also contribute to the scattering process considered here for finite $m_e$, but is additionally suppressed by the small $m_e/M\sim10^{-3}$ ratio and is neglected. In the kinematics relevant for atomic calculations $E\approx0$, this latter amplitude $\phi$ is the one of interest, while the former $\Phi$ vanishes when evaluated at $E=0$.}
\beqn
{\rm Re}\Phi(E,t)&=&\frac{2E}{\pi} {\cal{P}}\int_{E_\pi}^\infty
\frac{dE'}{E'^2-E^2}{\rm Im}\Phi(E',t).
\eeqn
\indent
The principal value integral can be done analytically by changing the order of integration,
%\beqn
%$\int_{E_\pi}^\infty dE' \int_{E_\pi}^{E'}d\omega=\int_{E_\pi}^{\infty}d\omega\int_{\omega}^\infty dE'$, 
%\eeqn
and I obtain,
%\begin{widetext}
\beqn
&&{\rm Re}\Phi(E,t)=\frac{-t}{4\pi^3}
\int_{E_\pi}^\infty\frac{d\omega}{\omega}\sigma_T(\omega)
\ln\left(\frac{4\omega_{cm}^2}{-t}\right)\label{eq:result}\\
&&\times\left[\left(1+\frac{\omega^2}{2E^2}\right)\ln\left|\frac{E+\omega}{E-\omega}\right|+\frac{\omega}{E}\ln\left|1-\frac{E^2}{\omega^2}\right|-\frac{\omega}{E}\right].\nn
\eeqn
%\end{widetext}
This is the master formula that is a more general result than that of Ref. \cite{Gorchtein:2006mq} where the high energy approximation for the cross section was made. %In the next section we present the numerical evaluation of this integral.
%\section{Results}
%\label{sec:results}

\begin{figure}[h]
{\includegraphics[width=7cm]{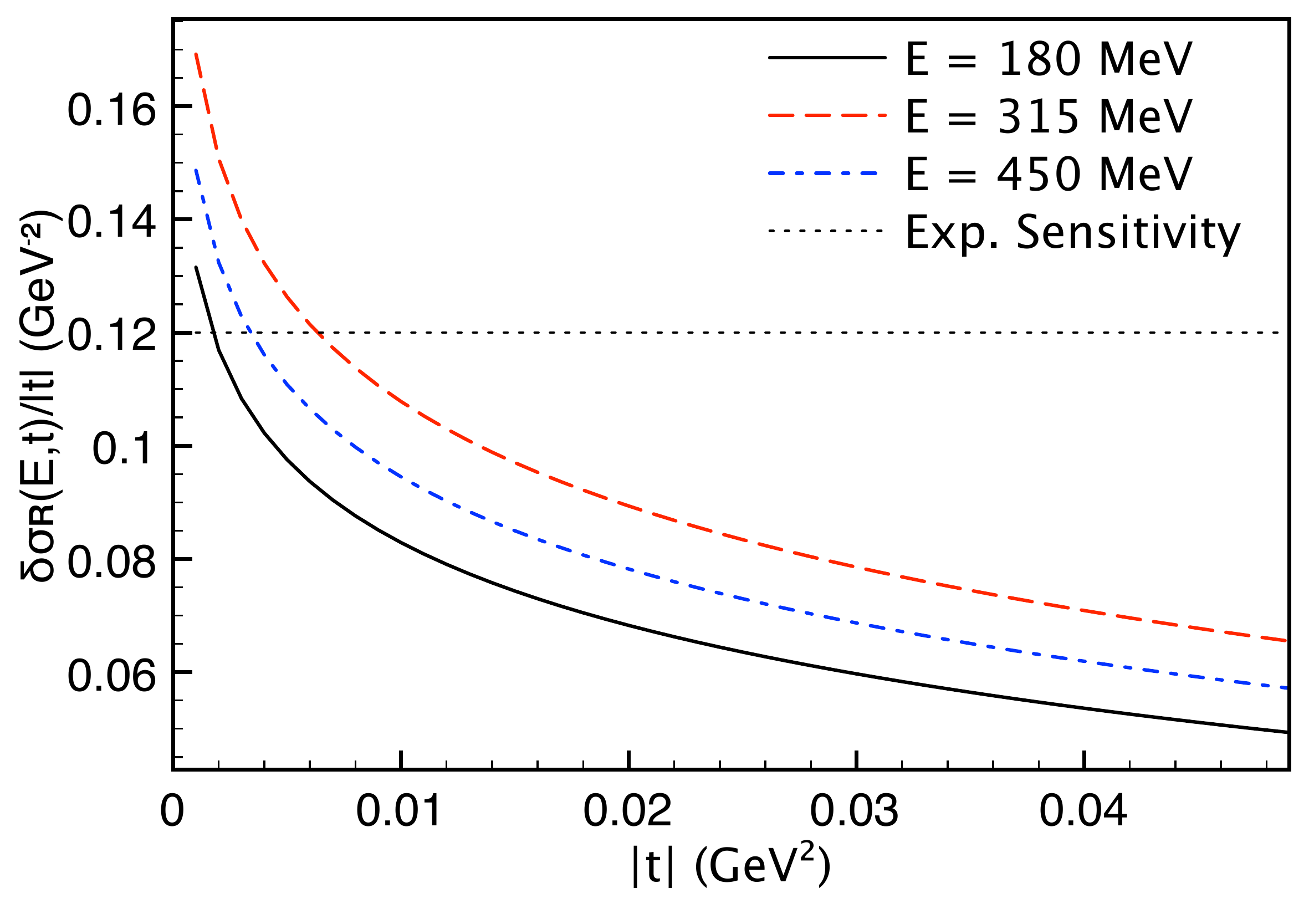}}
\vspace{-0.5cm}
\caption{(Color online) Results for the TPE effect on the reduced cross section $\delta \sigma_R(E,t)/|t|$ for the proton, as function of $|t|$ in GeV$^2$ for three values of the LAB beam energies:\\ 180 MeV (solid black curve), 315 MeV (long-dashed red curve), and 450 MeV (dot-dashed blue curve). The experimental sensitivity is shown by a thin dotted horizontal line.}
\label{fig:TPEresults-p}
%\vspace{-1cm}
\end{figure}

\begin{figure}[h]
{\includegraphics[width=7cm]{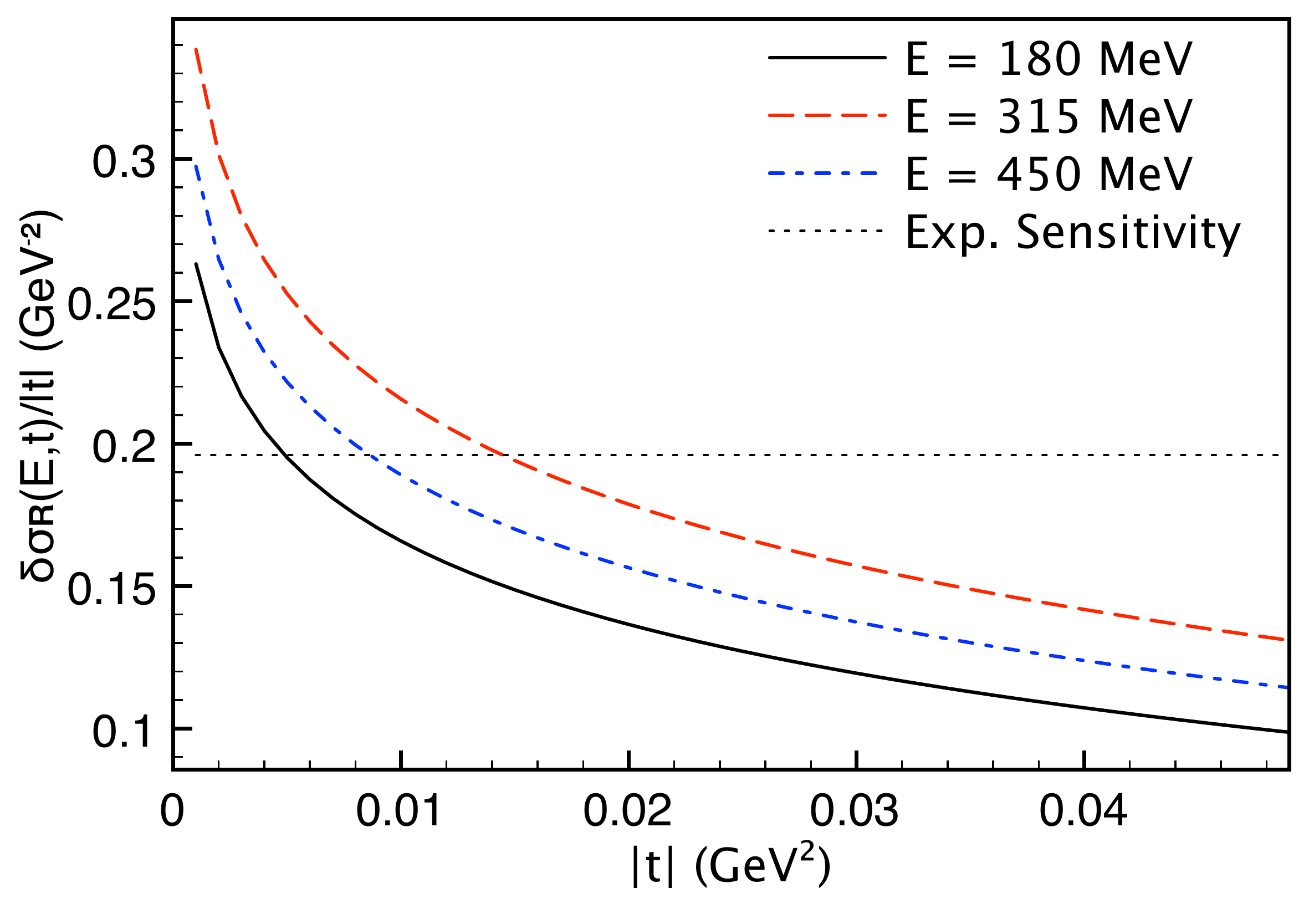}}
\vspace{-0.5cm}
\caption{%Results for the dispersive TPE effect on the reduced cross section $\delta \sigma_R(E,t)/|t|$ for the deuteron. 
(Color online) Same as in Fig. \ref{fig:TPEresults-p} for the deuteron.}
\label{fig:TPEresults-d}
\end{figure}

\begin{figure}[h]
{\includegraphics[width=7cm]{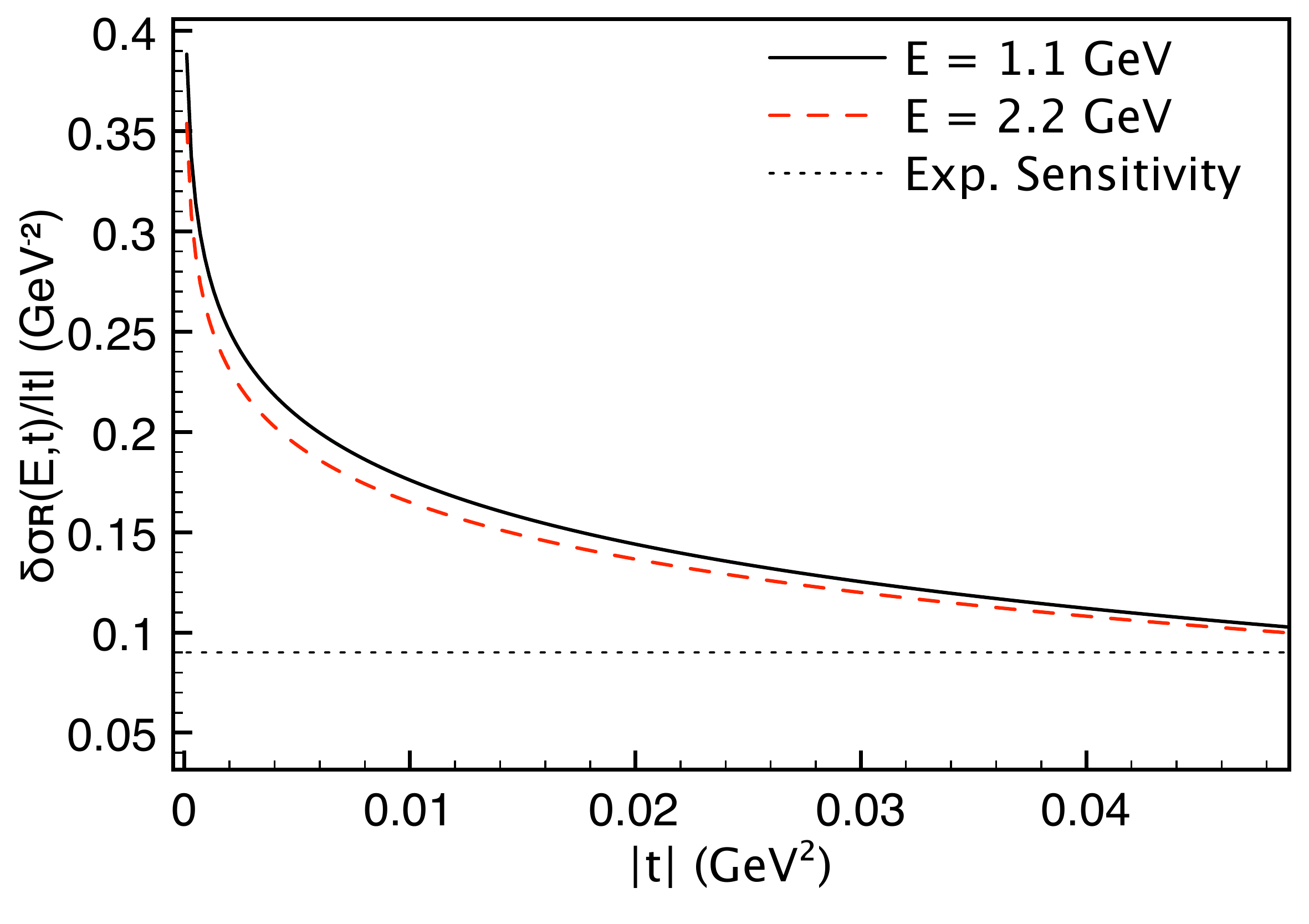}}
\vspace{-0.5cm}
\caption{
(Color online) Same as in Fig. \ref{fig:TPEresults-p} for the proton and for two values of the LAB beam energies: 1.1 GeV (solid black curve) and 2.2 GeV (long-dashed red curve) corresponding to the kinematics of the proposed JLab experiment \cite{JLab}.}
\label{fig:TPEresults-p2}
\end{figure}

The integral of Eq. (\ref{eq:result}) can be evaluated numerically using the phenomenological fit \cite{Gorchtein:2011xx}
 of the world data on real photoabsorption on the proton target \cite{PDG}. In Figs. \ref{fig:TPEresults-p}, \ref{fig:TPEresults-d}, \ref{fig:TPEresults-p2}, I present results for the quantity
 \beqn
\delta \sigma_R^{TPE} / |t| = 2\,{\rm Re}\Phi(E,t) / |t| ,
 \eeqn
that features the logarithmic behavior at low $|t|$, at three values of the electron beam energy relevant for the Mainz A1 experiment with the proton \cite{Bernauer:2010wm} and the deuteron \cite{Distler} target, the latter being currently under analysis. It is compared to the experimental sensitivity of the Mainz experiments that is obtained as 
\beqn
\delta \sigma_R^{R_{E},\,exp.} / |t| &=& -R_E^2/3 \left(1\pm \delta R_E/R_E\right).
\eeqn
For the proton, the experimental result of $R_E=0.879(8)$ fm translates into
\beqn
\delta \sigma_R^{R_{E},\,exp.} / |t| &=& -6.61(12)\,{\rm GeV}^{-2}.
\eeqn
This experimental sensitivity is compared in Fig \ref{fig:TPEresults-p} to the numerical evaluation of Eq. (\ref{eq:result}) in the kinematics of the A1 \@ Mainz experiment \cite{Bernauer:2010wm}
The energy dependence (difference between the solid, dashed and dash-dotted lines) reflects the energy dependence of the photoabsorption cross section around the $\Delta(1232)$ region. 
For the deuteron the projected precision of 0.25\% \cite{Distler} together with the recent global extraction of the deuteron radius from scattering and spectroscopy data $R_E^d=2.1424(21)$ fm \cite{Mohr:2012tt} leads to
\beqn
\delta \sigma_R^{R_{E},\,exp.} / |t| &=& -39.278(196)\,{\rm GeV}^{-2},
\eeqn
the uncertainty corresponding to a 0.25\% projected precision of the scattering experiment. A somewhat smaller value of 2.130(3) fm was extracted from the electron scattering data alone in Ref. \cite{SickTrautmann}; however, the difference is of no numerical importance for the analysis presented here. A comparison of this sensitivity to the TPE correction in the kinematics of A1@ Mainz experiment is displayed in Fig. \ref{fig:TPEresults-d}.
The result for higher energy relevant for the proposed JLab experiment \cite{JLab} at higher energies is shown in Fig. \ref{fig:TPEresults-p2}
As already mentioned, the TPE results in the leading logarithm approximation are model-independent modulo a {\it const.}$\,\cdot\,t$ offset  that translates into a constant in Figs.  \ref{fig:TPEresults-p}, \ref{fig:TPEresults-d}, \ref{fig:TPEresults-p2}. Therefore,  what really matters is not the absolute value of the TPE correction but rather the difference between the lowest and highest values of $t$ (i.e., nonlinearity). This nonlinearity is close to the experimental precision for the proton and the deuteron at moderate energies, as in the A1 @ Mainz kinematics, but is seen to be roughly three times the experimental sensitivity for the energy in the range of a few GeV and $|t|$ between $10^{-4}$ and $5\times 10^{-2}$ GeV$^2$ as in the proposed measurement at JLab. This suggests that the leading logarithm TPE correction has to be included in the experimental analyses that aim at extracting the charge radius from electron scattering with an accuracy below 1\%. It is seen that the inclusion of the TPE correction leads to a stronger $t$-dependence at low momentum transfer. Upon subtracting the positive-definite $|t|\ln(4E^2/|t|)$ correction from the experimental data, the extracted value of the charge radius will necessarily decrease. The subleading corrections $O(\alpha t)$, not included in this calculation can also affect the extracted value of the charge radius. The second term in Eq. (26) contributes $\sim1\%$ to the radius \cite{Lee:2014uia}. However, further corrections $O(\alpha t)$ may reduce this number as, e.g., the third term in Eq. (26) does.

The TPE effect for the deuteron is somewhat larger than for the proton in comparison with the respective experimental sensitivity. This can be understood by recalling that the total photoabsorption cross section for the deuteron is roughly twice that for the proton in the hadronic range. On the other hand, the quantity $R_E^2(\delta R_E/R_E)$ is only about 1.5 times larger for the deuteron giving a larger relative effect. Nuclear effects were neglected in this estimate. Moreover, the deuteron quasielastic break-up was not included: the derivation is based on treating $t$ as small compared to all other scales, an approximation that would not be valid if $t$ were to be compared to the characteristic scale $M B_d\approx2\times10^{-3}$ GeV$^2$, with the deuteron binding energy $B_d\approx2.224$ MeV. An exact calculation would be needed to account for the nuclear part of the photoexcitation of the deuteron. 

In summary, I have considered elastic electron-proton (deuteron) scattering at low momentum transfer $t$ and in the presence of the two-photon exchange (TPE). I calculated the TPE effect on the unpolarized cross section in the limit of low $t$. For the TPE effect with just the nucleon degrees of freedom inside the loop (elastic contribution), the leading behavior $\sim\sqrt{-t}$ is given by the model-independent Feshbach correction. For the TPE effect with inelastic states, the leading low-$t$ behavior is $t\ln t$, and the coefficient in front of this term is model-independent and given by a weighted integral over the total photoabsorption cross section. This integral was evaluated numerically using the recent parametrization of world total photoabsorption data on the proton and the deuteron, and the result was compared with the experimental accuracy in extracting the charge radius from electron scattering data at low $t$ in the kinematics of recent and upcoming experiments. I found that while at the beam energy of a few hundred MeV, as in A1@Mainz the nonlinearity introduced by the $t\ln t$ TPE correction is comparable to the experimental precision. At higher energies of 1-2 GeV corresponding to the experiment planned at  Jefferson Lab, this effect becomes about three times larger than the experimental precision for extracting the proton charge radius, and must be included. 

%\begin{acknowledgments}
%\vspace{0.5cm}
\acknowledgments{The author is grateful to M. Vanderhaeghen for pointing out that Eq. (34) was first derived in \cite{Brown:1970te} and applied to the $e^+/e^-$ cross section ratio at high energies, and to M. Distler, V. Pascalutsa, O. Tomalak, S. Karshenboim for useful discussions. This work was supported by the Deutsche Forschungsgemeinshaft DFG through the Collaborative Research Center ``The Low-Energy Frontier of the Standard Model" (SFB 1044) and the Cluster of Excellence ``Precision Physics, Fundamental Interactions and Structure of Matter" (PRISMA). }
%\end{acknowledgments}

\end{document}